\documentclass[fleqn,usenatbib]{mnras}
\usepackage{amsmath}
\usepackage{mathptmx}
\usepackage{txfonts}
\usepackage[T1]{fontenc}
\usepackage{ae,aecompl}
\usepackage{graphicx}
\usepackage{amssymb}
\usepackage[normalem]{ulem}
\usepackage{longtable}
\usepackage{multirow}
\usepackage{tablefootnote}

\newcommand{\hmpc}{{\, h^{-1}\, {\rm Mpc}}}
\newcommand{\changeR}[1]{\textcolor{red}}

\def\apj{ApJ}
\def\apjl{ApJL}
\def\apjs{ApJS}

\def\mnras{MNRAS}
\def\aap{A\&A}

\def\apjs{ApJS}
\def\apjl{ApJ Letters}

\title[Local environment of flat galaxies]{The local environment of flat galaxies }
\author[Sarkar, S., Banerjee, A., Makarov, D.] {Suman Sarkar$^{1}$ Arunima Banerjee $^{1}$\thanks{arunima@iisertirupati.ac.in}, Dimitry Makarov $^{2}$ \\
$^1$ Department of Physics, Indian Institute of Science Education and Research ( IISER ) Tirupati, Tirupati - 517507, India \\
$^2$ Special Astrophysical Observatory, Russian Academy of Sciences, 369167, Nizhnii Arkhyz, Russia}

\date{\today}

\pubyear{}

\begin{document}
\label{firstpage}
\pagerange{\pageref{firstpage}--\pageref{lastpage}}      
\maketitle

\begin{abstract}

The existence of flat or bulgeless galaxies poses a challenge to the hierarchical structure formation scenario advocated by modern cosmology. We determine the geometrical environment of a sample of $315$ flat galaxies from the Revised Flat Galaxy Catalog (RFGC) using `local dimension' $D$, which, on a given length scale, quantifies the dimension of the cosmic structure in which a galaxy is embedded. For galaxies residing in filaments, nodes 
and sheets, $D \sim 1$, $D \sim 1.5$ and $D \sim 2$ respectively; $D \sim 3$ represents field galaxies. We also determine the local dimensions of a sample of 15,622 non-flat galaxies identified in the Galaxy Zoo project from the Sloan Digital Sky Survey (SDSS). We find that the median values of $D$ for the flat and the non-flat galaxies are $2.2$ and $1.8$ respectively, implying that flat galaxies are located in a relatively sparser environment compared to non-flat galaxies; a Kolmogorov-Smirnov (KS) test indicates that their geometrical environments are different at $>99$\% confidence level. Further,  using a group finding algorithm, we study the local environment of a subset of $779$ flat galaxies with major-to-minor axes ratio $a/b >10$ identified as superthin galaxies. We find that the median clusterization index $k_{\rm{min}}$ for superthin flat galaxies $\sim$ $2.3$  while $\sim$ $1.7$ for other flat galaxies, confirming that the superthins reside in an under-dense environment compared to other flat galaxies at $> 98 \%$ confidence level. Our results may therefore have important implications for the formation and evolution models of flat galaxies in the universe.

\end{abstract}


\section{Introduction}
Galaxies are considered to be elementary units of the large scale structures of the universe. Various studies show that 
distribution of galaxies is not merely random, rather they form a complex network commonly called as the {\it cosmic web}. 
The cosmic web consists of huge voids that are surrounded by sheets or walls formed by galaxies. The intersection of these 
walls are linear string of galaxies called filaments. The intersection of filaments make nodes or clusters of galaxies. 
Earlier studies suggest that structures on cosmic scale exhibit asymmetric ellipsoidal collapse  \citep{lin65,zel70}. Systematic 
mass flow happens from voids to sheets, sheets to filaments and finally filaments to nodes \citep{rama15}. This makes 
the filaments denser than sheets and nodes denser than filaments. A galaxy in the center of a node or cluster will surely have a 
different environment compared to a galaxy embedded in a sheet. \\
   
\noindent It is  a well accepted fact now that the local environment plays a crucial role in regulating the physical 
properties of a galaxy. Intrinsic processes like star-formation, AGN activity and quenching affect the physical 
properties of a galaxy. However, there are several external processes like tidal interaction, galaxy-merging and ram-pressure 
stripping which can change the characteristics of a galaxy as well. All of these processes might have a direct or indirect link with 
the local environment. Several studies show that there is a strong relationship between the galaxy morphology and environment 
\citep{davis76,dressler80,guzzo97}. A number of studies also connect the color and stellar mass of galaxies with their 
local environment \citep{baldry07,bamford09,skibba09}. Observations suggest that red early-type galaxies are preferentially 
found in denser environments like clusters, whereas blue spirals are generally found in relatively rarer parts of the cosmic 
web. A recent study by \citet{pandey20} shows that apart from local density the color of a galaxy can also get affected by 
the type of cosmic structure it is part of. \citet{luparello15} suggests that properties like luminosity, stellar mass and 
color are much effected when galaxies reside in super-clusters where mass accretion dominates the local dynamical processes. 
Formation of galaxies is also strongly regulated by the underlying dark matter distribution. Because of the hierarchical halo formation 
and merging processes, the properties of a galaxy can also be dependent on the merger history of the dark matter halos they 
reside in. Due to the fact that different halos are assembled at different epochs, halos of different mass are expected to 
have different clustering strength. This assembly bias \citep{gao05,croton07} is the reason why massive galaxies are to be 
found in rich clusters. \citet{kauffmann04} suggests that on small scales specific star formation rate of a galaxy is deeply 
connected to the local density and the star formation history of the galaxy. \citet{zehavi02} has already shown that there 
is a clear difference in the correlation function of galaxies based on their surface brightness values. A study 
by \cite{mobasher03} in coma cluster shows that galaxies with low surface brightness (LSB) are found relatively lesser in 
cluster cores, indicating destruction of LSBs in dense environments. Some recent study by \citet{prole21,tanoglidis21} 
suggests that blue LSBs predominantly reside in low density environments and red LSBs are more clustered compared to the 
blue LSBs.\\

\noindent In this work we focus on the environment of flat (bulgeless) spiral galaxies. Formation of flat galaxies is yet 
an unsolved mystery in the current $\Lambda CDM$ paradigm. According to \citet{kormendy10} more than $65\%$ of the spiral 
galaxies in clusters have {\it classical-bulge}. These galaxies are considered to be early-type and they usually carry 
signature of major mergers. These galaxies usually have a higher metallicity and low star formation rate, as the gas 
content gets exhausted with time. This also makes them redder compared to the galaxies with a {\it pseudo-bulge}. 
\citet{kormendy10} also points out that $80 \%$ of the spirals in fields do not have a classical bulge. Hence, the bulge 
formation in galaxies with pseudo-bulge is more of a secular process. Another study in the COSMOS field 
\citep{grossi18} claims that star-forming bulgeless galaxies are mostly located in low density regions like fields and 
filaments. On the other hand \citet{governato10} shows through a hydro-dynamical simulation that the formation of bulgeless 
galaxies might be a result of strong outflow of low angular momentum gas from the center of the galaxy, which results in 
quenching the star formation near the central part of it. In such a case the galaxy is expected to undergo mergers and 
strong interactions being in denser environments. Hence, there is no clear consensus over the role of environmental in 
formation of flat galaxies. Characterisation of the local geometrical and physical environment of these galaxies can therefore be 
a useful tool to solve this puzzle.\\

\noindent A subset of the flat galaxies are the superthins. Superthin galaxies are edge-on galaxies with large axial ratios 
(major-to-minor axes ratio $a/b > 10$) with no discernible bulge component. They are late-type galaxies of the morphological 
type Scd/Sd and typically LSBs in nature. The term superthin was first coined by \citet{goad81} who made spectroscopic 
observations of four such galaxies. Later they were studied in general as a part of larger surveys or catalogues: For example, 
the Flat Galaxy Catalogue (FGC) \citet{kara93}, the Revised Flat Galaxy Catalogue (RFGC) of, and more generally, studies 
of edge-on galaxies \citep{dalcanton02, kregel05, kautsch06, comeron11, bizyaev14}. Being rich in neutral hydrogen gas (HI), 
superthin galaxies have also been studied as part of HI 21 cm survey of flat galaxies \citep{matthews98, matthews00, obrien10}. 
The origin as well as survival of superthin stellar discs continues to be a puzzle. The disc vertical structure is regulated 
by a balance between the components of the net gravitational field and the stellar velocity dispersion in the vertical 
direction. Although the discs are primarily rotation supported in the plane, they are supported by velocity dispersion in the 
vertical direction. Therefore a superthin stellar disc implies a low value of the vertical velocity dispersion 
of the stars i.e., an ultra-cold stellar disc  \citep{aditya21} and/or a high value of the net gravitational field in the vertical direction.
\citet{jadhav19} showed that superthin galaxies have characteristically larger disc sizes compared to other ordinary flat 
galaxies with equal specific angular momentum, which results in their large planer-to-vertical axes ratio.\\

\noindent The net gravitational potential of galaxies can be mapped by mass modeling studies using mainly HI rotation curves 
and/or HI vertical scale height data obtained from HI 21cm radio observations \citep{banerjee08, banerjee10, banerjee17, kurapati18}. 
In fact, \citet{banerjee13} showed that the superthin stellar disc in UGC7321 may be attributed to a dense and compact dark matter 
halo. Due to their edge-on orientation, it is not possible to have a direct measure of the vertical velocity dispersion of superthin 
discs. However, since disc heating is primarily caused by dynamical interactions with nearby galaxies in the local environment like 
major and minor mergers, studying the local environment of superthin galaxies may solve the puzzle of the origin and survival 
of these ultra-cold, razor-thin stellar discs.\\

\noindent In this paper, we study in detail the local environment of the flat galaxies and the superthin galaxies as a subset. 
We use the Revised Flat Galaxy Catalog (RFGC) data to first determine the local dimension and local density estimators to characterize 
the local environment of the flat galaxies including the superthins. We compare the results with that of the non-flat spiral 
galaxies as identified in the Galaxy Zoo project to investigate if the flat galaxies reside in any special type of geometrical environments compared to other non-flat spirals. We then employ the group finding method of \citet{makarov11} to compare and contrast the local environments of superthin flat galaxies and other flat galaxies (See also, \citet{kara16}). \\

\noindent The rest of the paper is organized as follows: In \S 2, we discuss the data, in \S 3 the method of analysis, in \S 4 
the results followed by discussion and conclusion in \S 5 and 6 respectively.

\section{DATA}
\subsection{RFGC data : Flat spirals }
The revised flat galaxy catalogue (RFGC) \citep{{kara99}} is the amended version of the flat galaxy catalogue (FGC)\citep {kara93}. 
It provides detailed description of $4236$ galaxies observed in the nearby universe having axial ratio of $a/b \geq 7$ in their blue image. 
Unlike its predecessor, it also provides the red diameters of these galaxies. The coordinate measurements is also improved upto 
$\sim 3 ^{\prime\prime}$ in the revised version of the catalogue. We identify the galaxies in RFGC and use a proper matching technique of 
galaxy IDs to find them in the twelfth data release (DR12) \citep{alam15} of Sloan Digital Sky survey (SDSS) database to get the 
information about their redshifts. The Vizier X-match server \footnote{http://cdsxmatch.u-strasbg.fr/} is used to uniquely find galaxies 
across the two different databases. The matched data consists of $1394$ flat galaxies, majority of them lying within the redshift range 
$0.01 \leq z \leq 0.09 $. Within the given redshift range and angular span specified in $\S 2.2$ for the SDSS DR16 spectroscopic sample
(used for mapping the neighbourhood) we are left with $315$ flat galaxies. Using the information of redshift and angular coordinates, a 3D distribution of flat galaxies is mapped. The mapped cartesian coordinates of these bulge-less galaxies are used to carry the analysis 
discussed in $\S 3.1$. For comoving distance calculation from redshifts, we use the $\Lambda$CDM model with $\Omega_{m0}=0.315$, $\Omega_{\Lambda0}=0.685$ and $h=0.674$ \citep{planck18}.

\subsection{SDSS DR16 Spectroscopic data}
The Sloan Digital Sky Survey (SDSS) \citep{york00} is one of the leading redshift surveys to date. SDSS provides spectral sky coverage 
of $9,376$ square degrees targeting around 3 million galaxies in its third phase. We extract data from the sixteenth data release (DR16) 
\citep{ahumada19} of SDSS.  A SQL query in the CASjobs \footnote{https://skyserver.sdss.org/casjobs/} is used to extract data from SDSS 
public database. Due to the non-uniform sky coverage of SDSS, a contiguous region is identified in the northern galactic hemisphere 
between right ascension $125^\circ \leq \alpha \leq 235^\circ$ and declination $0^\circ \leq \delta \leq 60^\circ$. We prepare a 
magnitude limited samples with {\it r}-band Petrosian absolute magnitude limit $M_r \leq -20.5$ and consider the redshift range 
$ 0.01 \leq z \leq 0.09$, which provides us $107290$ galaxies with a mean number density of  $1.05 \times 10^{-2} \, h^3 {\rm Mpc}^{-3}$. 
We prepare a 3D distribution of the SDSS spectroscopic galaxies to map the neighbourhood of our sample galaxies. We use this 
neighbourhood to characterize the local environment of the flat galaxies in RFGC catalog as well as the non-flat galaxies from Galaxy zoo. 

\subsection{Galaxy Zoo2 : Non-flat spirals}
We use the {\it Zoo2Specz} table from the database of the twelfth data release (DR12) of SDSS. {\it Zoo2Specz} contains morphological
classification of galaxies provided by the {\it Galaxy zoo2} \citep{willet13} project. Galaxy zoo 
\footnote{https://www.zooniverse.org/projects/zookeeper/galaxy-zoo/} \citep{lintott08} is a citizen science 
program where morphological classification of galaxies is done through visual inspection followed by a rigorous likelihood analysis. 
Galaxy zoo2 is the second phase of Galaxy zoo where a much detailed classification is made for a subset of galaxies from Galaxy zoo. 
We use a structured query to identify the spiral galaxies which have a obvious bulge or a noticeable bulge and discard those having 
no bulge. We select these galaxies within our selected red-shift range $0.1 \leq z \leq 0.09$ and also restrict them between the angular 
span ($125^\circ \leq \alpha \leq 235^\circ$, $0^\circ \leq \delta \leq 60^\circ$). We get a sample of $15622$ non-flat galaxies in the 
selected region.\\

\begin{figure*}
\resizebox{8 cm}{!}{\rotatebox{0}{\includegraphics{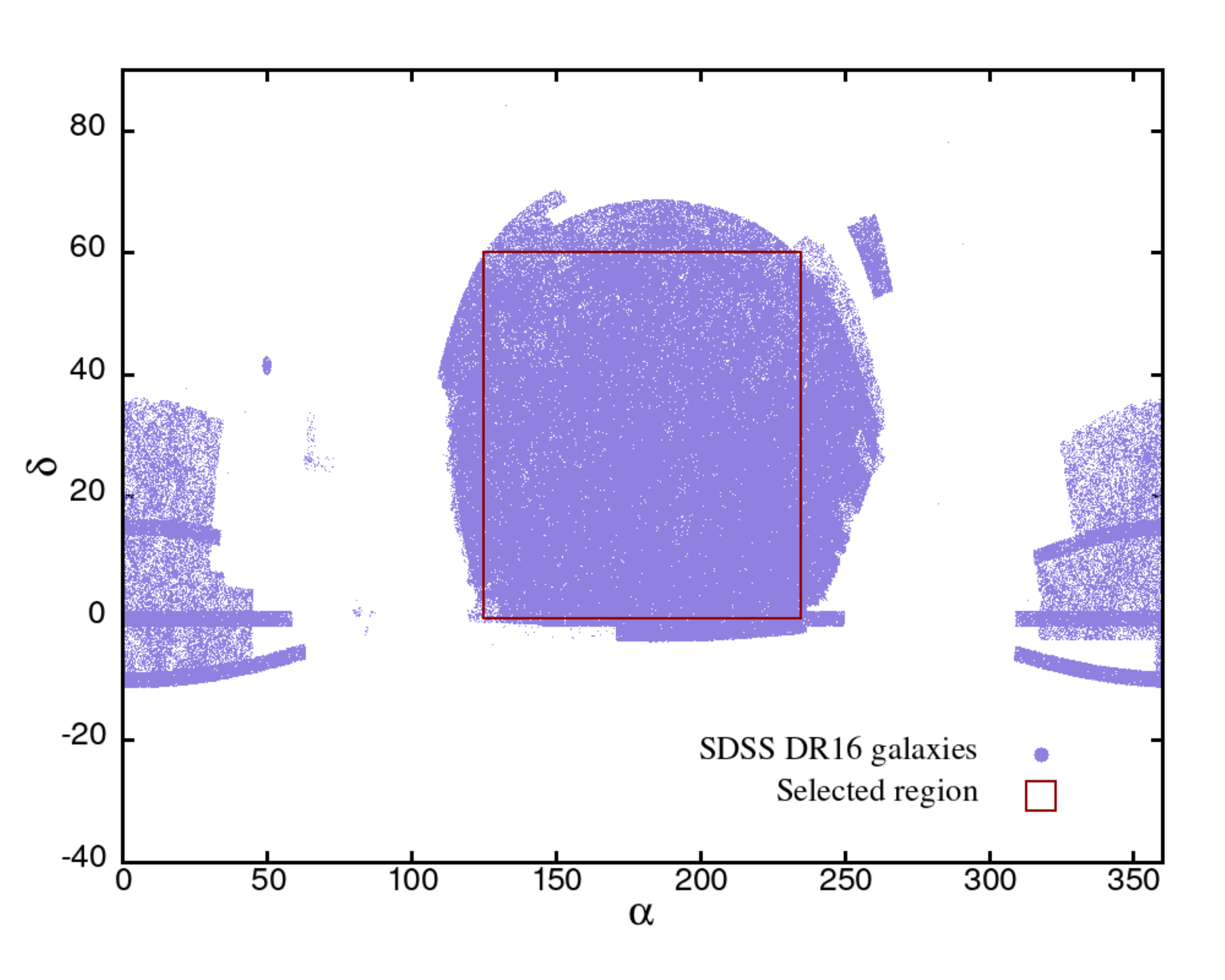}}} \hspace{1 cm}
\resizebox{8 cm}{!}{\rotatebox{0}{\includegraphics{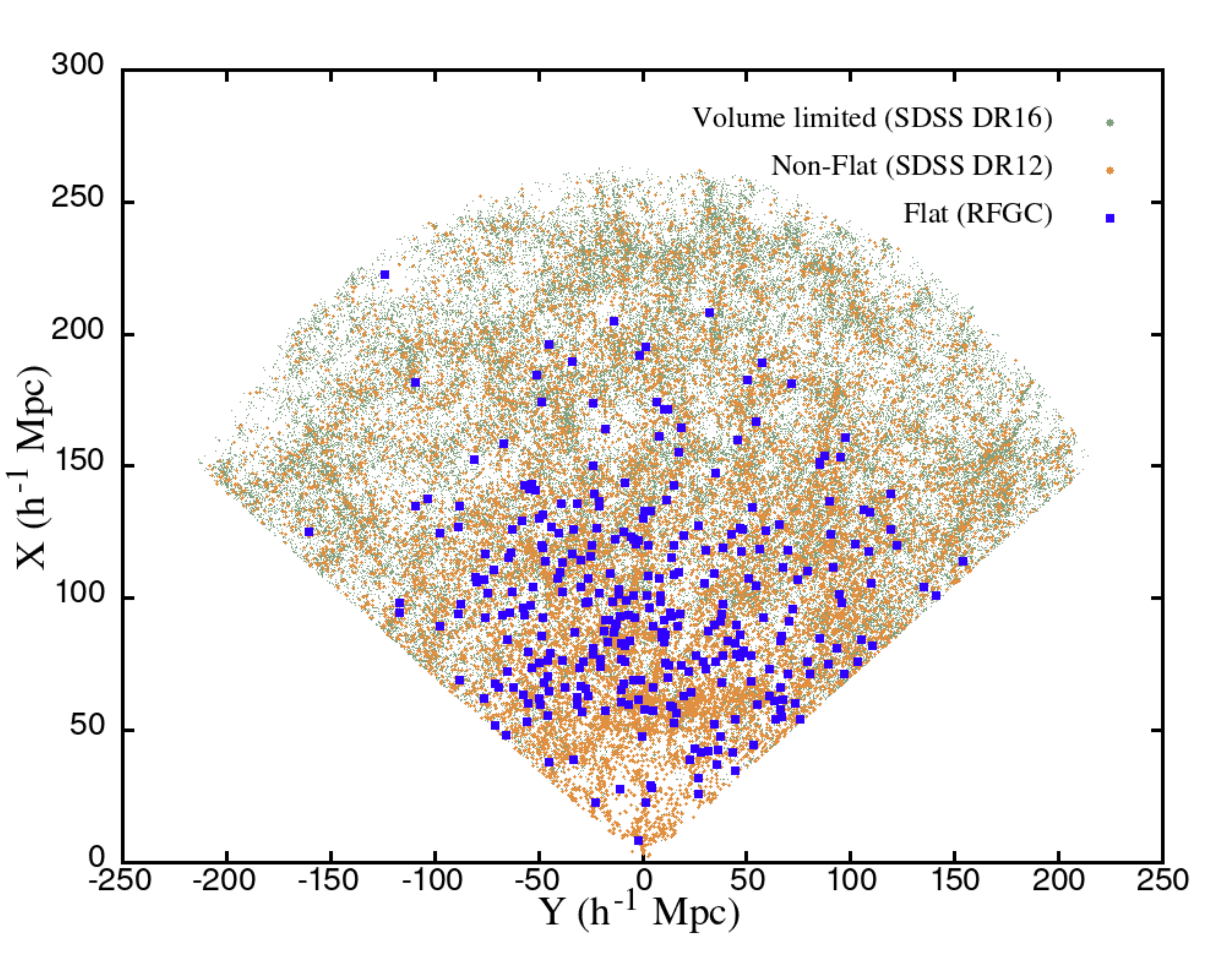}}} \\ 
\label{fig:th_data}
\caption{The left panel of this figure shows the contiguous region considered in SDSS sky coverage. The right panel shows the  
distribution of flat galaxies and non-flat galaxies along with the neighbouring SDSS DR16 galaxies within the coverage shown on left, 
all projected on the X-Y plane.}
\end{figure*}

\subsection{RFGC data for group finding}

Our sample for the above analysis consists of $2,762$ galaxies from the RFGC \citep{kara03}, 
a catalogue of $4,236$ thin edge-on spiral galaxies and covering the whole sky, our sample size being constrained by the dearth of 
information pertaining to galactic environment for some of the RFGC galaxies. The RFGC galaxies excluded from our sample were either 
(i) at too high a redshift (ii) located at the zone of avoidance ($-10^{\circ} < galactic\,\,\, longitude < 10^{\circ}$) or (iii) 
with no measured value of  $V_{LG}$ available. For the group finding part we choose the galaxies with major-to-minor axis ratio 
\textbf{$a/b > 10$}. Among the $2,762$ RFGC galaxies in our sample, only $779$ are superthins. while the rest are non-superthin galaxies. 
Therefore superthin galaxies constitute  ($\sim 28 \%$) of our sample.

\section{Method of analysis}
\subsection{Geometrical environment of flat galaxies and connection with local density}
We aim to investigate if the flat galaxies reside preferentially in a sparser environment compared to non-flat galaxies by determining the local dimension of the these galaxies \citep{sarkar09}. We obtain the 
3D spatial map of the flat galaxies and non-flat galaxies using their red-shifts. We then use the volume limited sample, extracted from 
SDSS DR16 spectroscopic data to map the neighbouring galaxy distribution for each of the flat and non-flat galaxies.
Constructing a co-moving sphere of radius $R$ around each flat galaxy, we count the number of neighbouring DR16 
galaxies in that sphere. \\

\noindent If within the given span $ R_1 \le  R \le R_2$ a galaxy's local geometric environment does not change then in principle the relation
\begin{eqnarray}
 N(< R)= A R^{D}
\label{eq:ld1}
\end{eqnarray}
has to hold \citep{sarkar09}. Here $D$ is the local dimension of the cosmic web around that galaxy and $A$ serves as a proportionality 
constant. One would expect the values of $D$ to be 1, 2 and 3 respectively for galaxies in filaments, sheets and fields or volume filling 
structures \citep{sarkar19}. \\

\noindent Using logarithm on both sides of \autoref{eq:ld1} we get
\begin{eqnarray}
\log N = \log A + D\log R.
\label{eq:ld2}
\end{eqnarray}

\noindent \autoref{eq:ld2} represents a straight line where $D$ indicates the slope of the line and $\log{A}$ is the intercept. For different values 
of $R$ we find different values of $(\log{N},\log{R})$. The local dimension corresponding to a flat galaxy is obtained by performing a least square fit 
for all the values of $(\log N,\log R)$ for different $R$. We vary $R$ with a uniform increment of $\Delta R=0.1 \times (R_2-R_1)$. 
Figure 2 shows the goodness of fit for three different values of $D$.\\

\noindent The definition of local dimension given here is somewhat similar to that of {\it fractal dimension} \citep{mandelbrot83}. However, 
the process of evaluation of these two identities and the purpose are different from each other. The fractal dimension is used to investigate 
the self similarities in the structural form of the test subject on different length scales. Whereas, the local dimension on a given length 
scale quantifies the local geometric environment of a galaxy embedded in a larger structure.\\

\noindent We use chi-square minimisation technique to find those galaxies which have a steady value of $D$ within the probed length scale range. 
We set a stringent criteria $\frac{\chi^2}{\nu} \leq 0.25$ to ensure the goodness of the fit. Here, $\nu$ is the degree of freedom.
The chi-square value is calculated as 

\begin{eqnarray}
\chi^{2} = \sum_{\nu+1}{\frac{\left( \log{N} \right)^2 - \left( \log{ \left[ \bar{A}R^{\bar{D} }\right]} \right)^2} {\log \bar{A}R^{\bar{D}}}}.
\label{eq:chisqr}
\end{eqnarray}

\noindent Here $\bar{A}$ and $\bar{D}$ are respectively the best-fitting values of the intercept and slope obtained from the least square fit.\\

\begin{figure*}
\resizebox{9 cm}{!}{\rotatebox{0}{\includegraphics{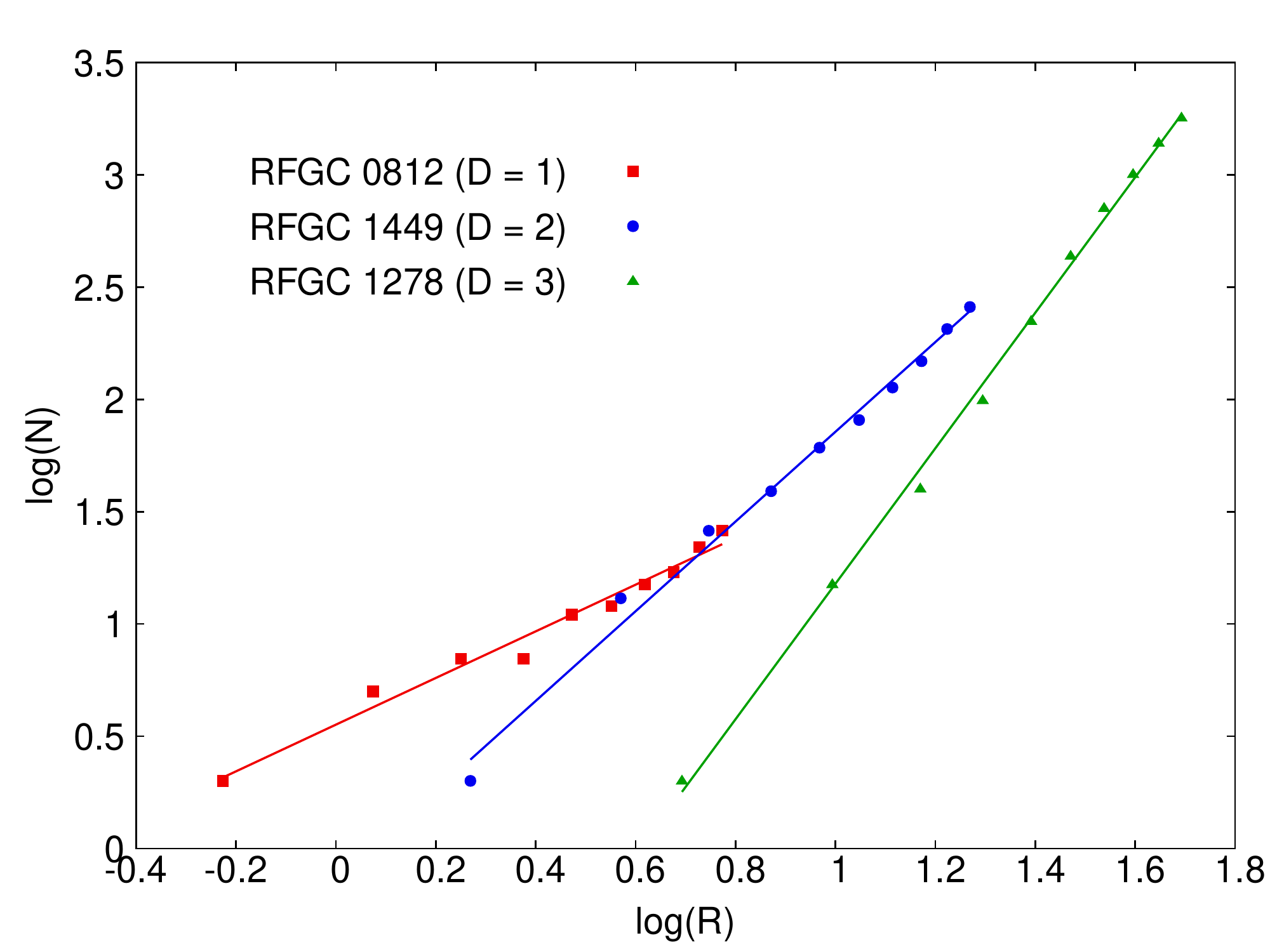}}}
\label{fig:dfit}
\caption{This figure shows the goodness of fit for the least square fitting between the number of galaxies $N$ within a co-moving radius $R$
centered on a sample flat galaxy. Fitting for three arbitrary flat galaxies with three different values of $D$ are shown for comparison.}
\end{figure*}

\noindent According to \citet{sarkar19} galaxies with $D \sim 1$ and $D \sim 2$ reside in filaments and sheets respectively. Galaxies with 
$D \sim 3 $ are those who are residing in a locally isotropic environment. For a dense galaxy sample very small values of $R_2$ for $D \sim 3$ 
galaxies can lead to galaxy clusters but for a larger value of $R_2$ they are mostly field galaxies. \\

\noindent In this work we have chosen not to use a fixed value of $R_2$, rather it is decided by the local density of the galaxy in question. In all 
the cases we choose $R_1$ to be the distance to the $2^{nd}$ nearest neighbour and set $R_2$ as

\begin{eqnarray}
R_2={\cal R}_{n} \times \exp \left[ \left( \frac{\eta}{\rho_m} \right)^{-\frac{1}{3}}\right]
\label{eq:R2form}
\end{eqnarray}

\noindent Here ${\cal R}_{n}$ is the distance to the $n^{th}$ nearest neighbour, ${\rho_m}$ is the mean number density of the volume 
limited sample prepared from SDSS DR16. $\eta$ here is the {\it local density} of the galaxy. The local density \citep{sarkar21} or 
core density \citep{casertano85, verley07} of a galaxy can be written as,

\begin{eqnarray}
\eta = \frac{3(n-1)}{4 \pi {\cal R}_{n}^3}
\label{eq:lden}
\end{eqnarray}

\noindent $R_2$ is taken in the form of \autoref{eq:R2form} to assign additional weightage on $R2$ for galaxies in relatively under-dense regions. This helps to properly segregate the galaxies in under-dense regions from the ones lying in over-dense environments. The local density ($\eta$) and the local dimension ($D$) both are functions of $n$ and in this case are coupled. A very large value of $n$ would lead 
to the scale of homogeneity. For this work, we have chosen the value $n=25$, which leads to ${\cal R}_{25} \sim 8 \hmpc$ and 
$R_2 \sim 22 \hmpc$ for the case where $\eta= \rho_m = 1.05 \times 10^{-2} h^3 {\rm Mpc}^{-3}$ is the mean number density of the 
background distribution. However, this choice is more of an empirical one, where the fitting is found to be optimum.\\

\noindent To statistically test if there is a difference between the local environment of the flat galaxies compared to the 
non-flat galaxies, we perform a two-sample Kolmogorov-Smirnov (K-S) test for the distributions of local density and local dimension.
The supremum difference between two cumulative probability distributions for any variable $x$ is obtained through K-S test as
\begin{eqnarray}
{\cal D}_{KS} (x) & = &  sup \, \, \{ \,\, | f_{m}(x)-g_{m}(x) | \,\, \}
\label{eq:Dks}
\end{eqnarray}

\noindent where $f_{m}(x)$ and $g_{m}(x)$ are the cumulative probability of variable $x$ for the two samples respectively, at $m^{th}$ 
checkpoint such that $\sum{f_{m}(x)}=\sum{g_{m}(x)}=1$. Here $sup$ is the operator that finds the supremum of all the obtained differences.\\

\noindent We find the corresponding probability ($p$) for accepting the null hypothesis that assumes the two distributions to be same. 
For an exact match $p$ will be 1, whereas the distributions are completely different if $p = 0$. In the second case we reject 
the null hypothesis. confidence level for rejection is $\alpha=(1-p)$. 

\noindent The value of $\alpha$ can be obtained from ${\cal D}_{KS}$ from the equation  
\begin{eqnarray}
\alpha= 1 - 2\sum_{i=1}^{\infty} {(-1)^{i-1} \exp \left[ -2i^2 {\cal D}_{KS}^2 \left( \frac{N_1 N_2}{N_1+N_2} \right)  \right] }
\label{eq:pks}
\end{eqnarray}
\noindent Here $N_1$ \& $N_2$ are number of data points in the first and second distribution respectively. If we set our critical 
$p$-value for rejection of the null hypothesis as $p=0.01$ then it will lead to a confidence level of $99 \%$. We can conclude the 
distributions are different when the $p$ value is smaller than the critical value. 


\subsection{Study of the environment of superthins using the group finding algorithm}

\noindent We determine the local environment of our sample galaxies by employing the {\it group finding algorithm} described in 
\citet{makarov11}. According to it, the bounded galaxy pairs are identified by imposing the constraints

\begin{itemize}
\item  the total energy is negative i.e.,
\begin{eqnarray}
\frac{V_{12,r}^2}{2GM_{12}}  <1. 
\label{eq:2.1}
\end{eqnarray}

\noindent where $V_{12,r}$ is the relative velocity of the galaxies projected along the line of sight and $M_{12}$ the total mass of the pair, and

\item the pair is within a zero-velocity surface i.e.,
\begin{eqnarray}
\frac{\pi H_0^2 R_{\perp}^3}{8GM_{12}}  <1. 
\label{eq:2.2}
\end{eqnarray}

\end{itemize}

\noindent where $H_0$ is the Hubble constant. If a galaxy is found to be paired up with multiple other galaxies, then the pair in which it 
has the most massive companion is considered. Next all the pairs with a common component is taken to constitute a group. Then, for each group, 
all the member galaxies are replaced by a single fake galaxy with a luminosity equal to the total luminosity of the group and a red-shift equal to the mean red-shift. Finally, the above steps are iterated until no new groups are detected. The galaxy masses are determined from their 
integrated luminosity in the near-infrared $K_s$ band,assuming they have the same mass-to-light ratio

\begin{eqnarray}
M/L_K = \kappa (M_{\odot}/L_{\odot}) 
\label{eq:2.3}
\end{eqnarray}

\noindent where $\kappa$ is taken to be equal to 6 at which the structure and the known virial masses of the groups is best reproduced \cite{kara11}.\\

\noindent The clusterization index $\kappa_{min}$ of a galaxy is defined as $\kappa=6 \times \kappa_{min}$ such that
\begin{eqnarray}
\frac{V_{12,r}^2}{2GM_{12}}  =1. 
\label{eq:2.4}
\end{eqnarray}

and 
\begin{eqnarray}
\frac{\pi H_0^2 R_{\perp}^3}{8GM_{12}}  =1. \nonumber
\label{eq:2.5}
\end{eqnarray}

\noindent has a sense of the isolation index $\Pi$ as defined in \citet{kara11}. $\kappa_{min}$ reflects the sparsity of the local environment. 
A higher value of $\kappa_{min}$ indicates a sparser environment and vice-versa. \\

\noindent {\bf The sub-samples:} A galaxy is defined to be isolated if $\kappa_{min} > 1$. Similarly, a galaxy is taken to be in a group 
environment if $\kappa_{min} < 1$. The galaxies in a group are further subdivided into the root galaxy which is the main member of the group 
and the rest are considered to be member galaxies. The definition of $\kappa_{min}$ is dependent on the status of the galaxy. For an isolated 
galaxy, $\kappa_{min} = min \left( \kappa_{root}, \kappa_{comp} \right) $ where $\kappa_{root}$ is the $\kappa_{min}$ of the galaxy calculated 
with respect to its possible closest satellite and $k_{comp}$ with respect to the most probable massive group of which it could be a member 
galaxy. For a root galaxy, $k_{min}$ is calculated with respect to the most important satellite, while for a member galaxy, it is the 
$\kappa_{min}$ at the moment of clusterization i.e., when it was included in the group. Besides, for galaxies in group environment, $\kappa_{root}$ 
is the $k_{min}$ calculated with respect to the most nearby satellite with the associated group taken as a single object. Similarly, 
$\kappa_{comp}$ is the $\kappa_{min}$ calculated with respect to more massive group, with the host galaxy group taken as a single object.

\section{Results}

\begin{figure*}
\resizebox{7.1 cm}{!}{\rotatebox{0}{\includegraphics{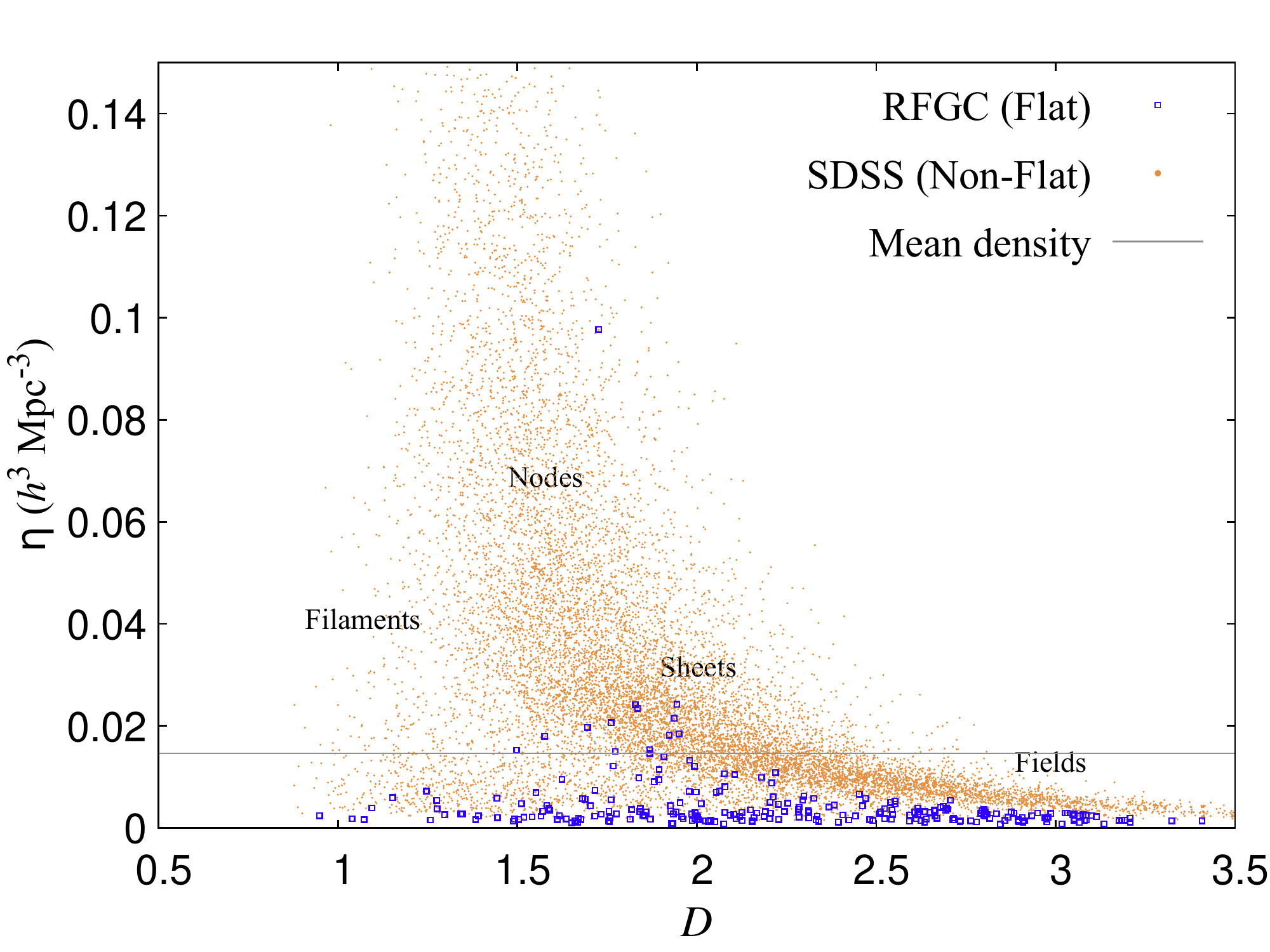}}} \hspace{0.3 cm}
\resizebox{7.1 cm}{!}{\rotatebox{0}{\includegraphics{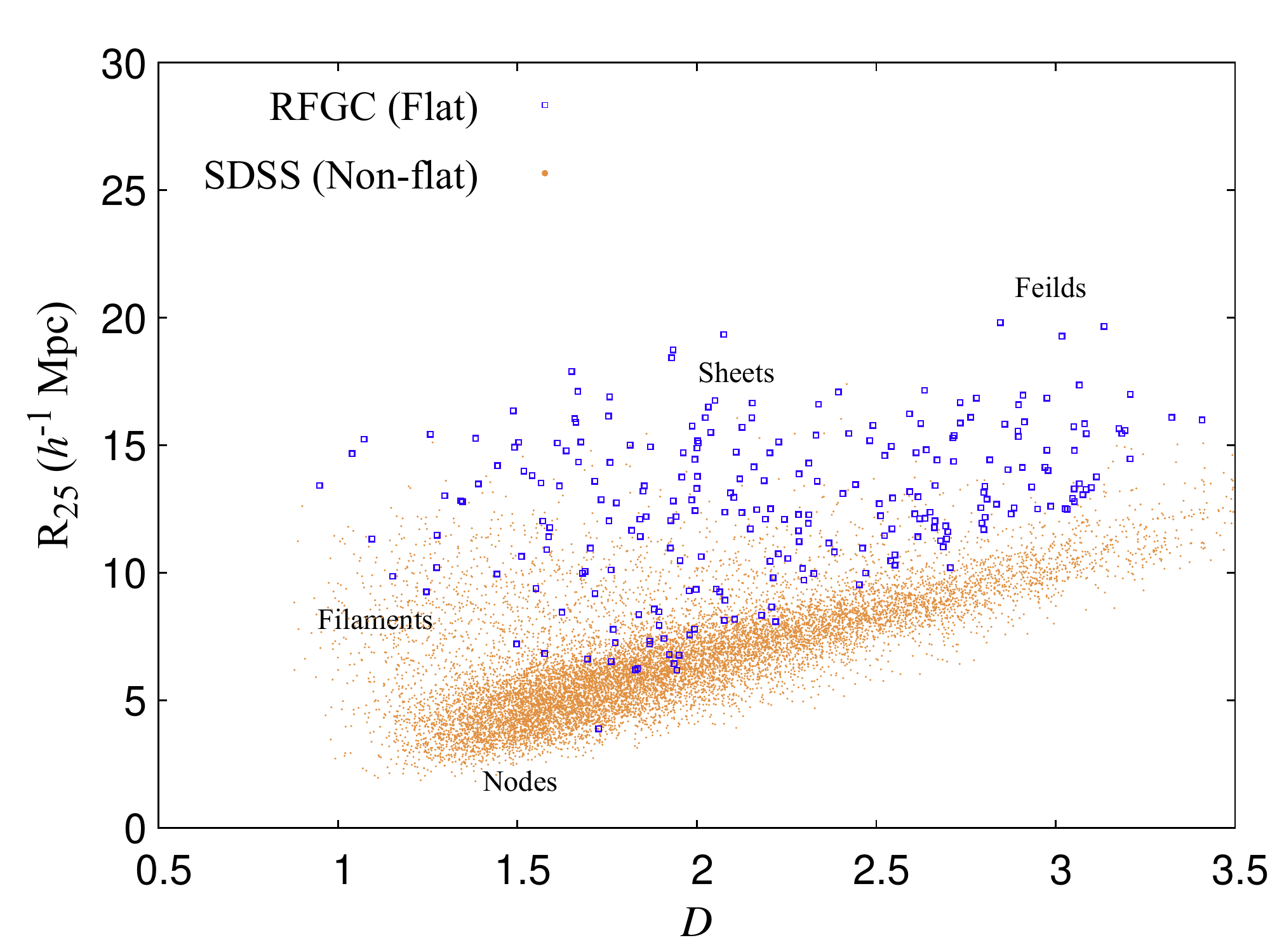}}} \\
\resizebox{7 cm}{!}{\rotatebox{0}{\includegraphics{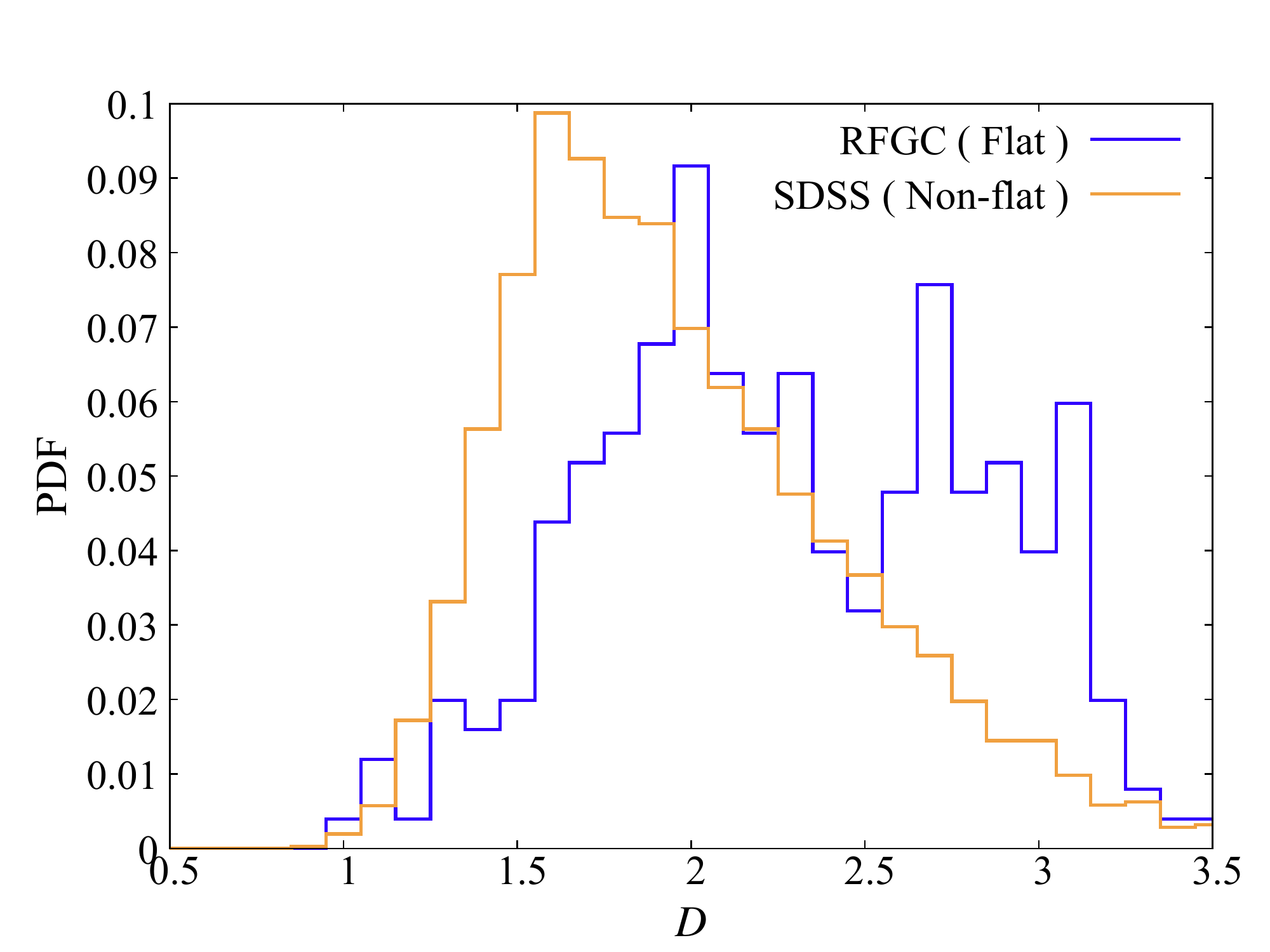}}} \hspace{0.5 cm}
\resizebox{7 cm}{!}{\rotatebox{0}{\includegraphics{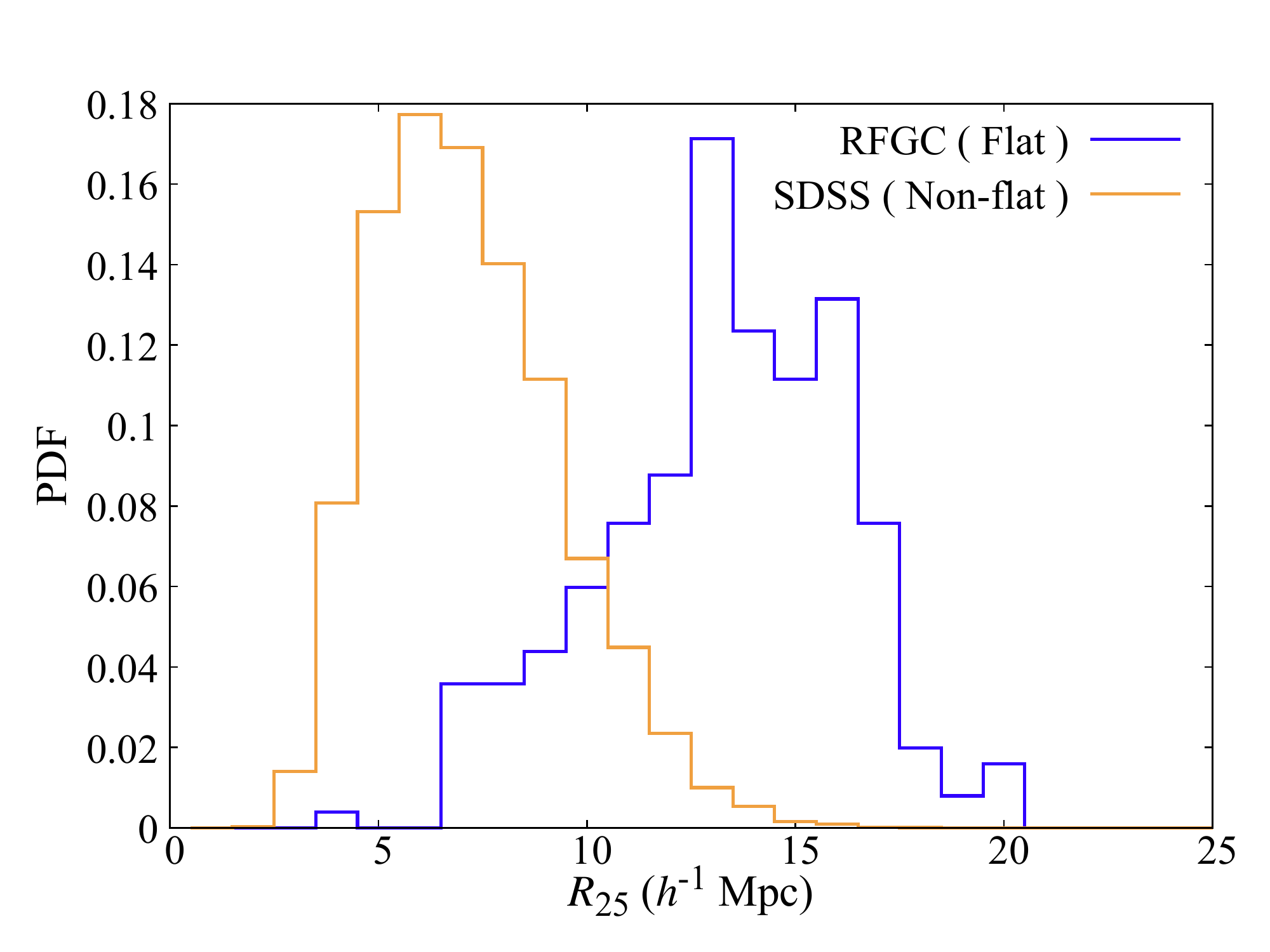}}} \\
\resizebox{7 cm}{!}{\rotatebox{0}{\includegraphics{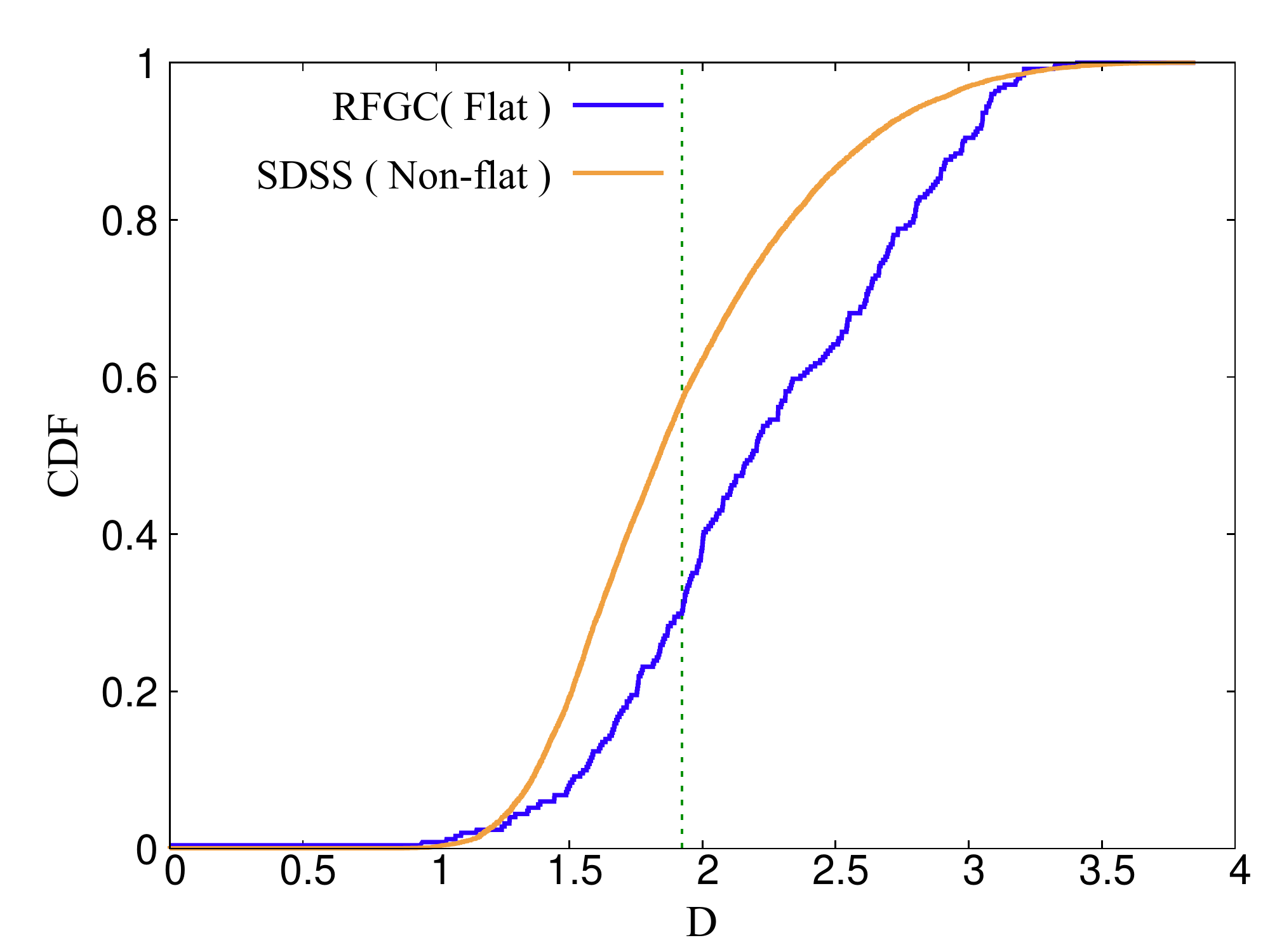}}} \hspace{0.5 cm}
\resizebox{7 cm}{!}{\rotatebox{0}{\includegraphics{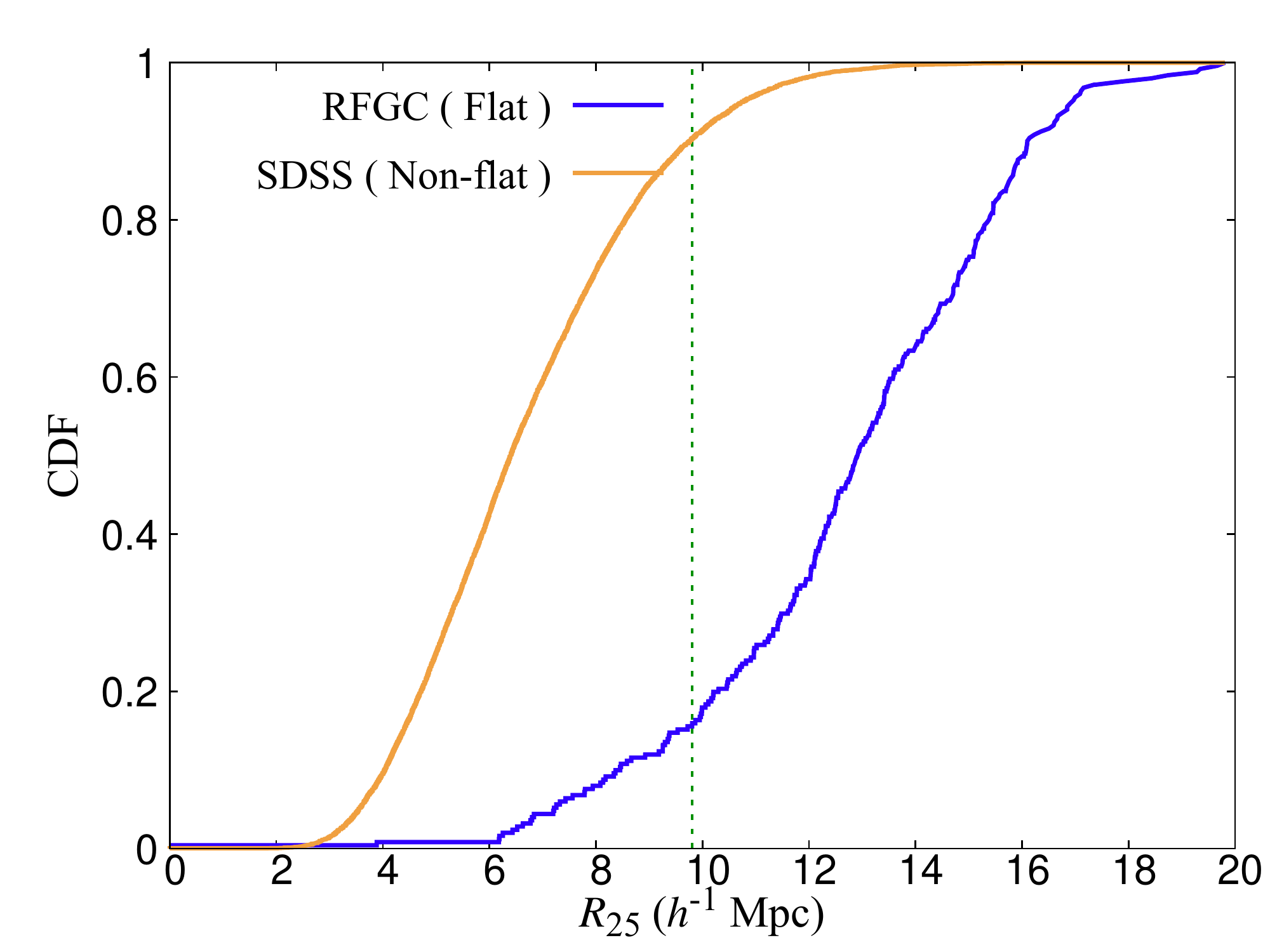}}} \\
\caption{The top left panel shows how the flat galaxies are distributed in the $D$-$\eta$ space along with the non-flat galaxies. 
The top right panel shows the same galaxies on the $D$-${\cal R}_{25}$ plane. The two middle-panels from left to right show the 
probability distribution functions for $D$ and ${\cal R}_{25}$ for the flat and non-flat galaxies. The cumulative distribution 
function is shown in the same way in two bottom panels. The dotted green lines in these plots show the point of supremum difference.}
\label{fig:ldrn}
\end{figure*}

\begin{table*}{}
\caption{Results: Geometrical environment of flat and non-flat galaxies}
\label{tab:1.1}
\begin{tabular}{lccc}
\hline
&$\eta$ & $D$ & ${\cal R}_{25}$ \\
\hline
Median ( flats ) & 0.003 $h^3 {\rm Mpc}^{-3}$ & 2.2 & 12.9 $h^{-1} {\rm Mpc}$ \\
Median ( non-flats) & 0.02 $h^3 {\rm Mpc}^{-3}$ & 1.8 & 6.4 $h^{-1} {\rm Mpc}$ \\   
${\cal D}_{KS}$ & $0.8$ & $0.3$ & $0.7$ \\
$p$ & $ 0$ & $1.98 \times 10^{-16}$ & $0$ \\
\hline
\end{tabular}
\begin {itemize}
\item [] \scriptsize{ $\eta$, $D$ and ${\cal R}_{25}$ are the local density, local dimension and the mean intergalactic separation resepctively. $\cal{D}_{KS}$ is the supremum distance in K-S test, $p$ the corresponding probability.}
\end{itemize}
\end{table*}

\subsection{Local dimension and local density of superthins}
The top left panel of \autoref{fig:ldrn} shows the distribution of superthins in the $D-\eta$ plane. The orange dots correspond 
to the non-flat galaxies whereas the blue points represent the flat galaxies. We find that majority of the non-flats reside in 
filaments or nodes ($D < 1.5$). However, the majority of flat galaxies are found to be lying in the sheets or fields ($D>2$). 
Although, some of them are found to be residing in the midst of filaments as well. Here we have to remember that all the galaxies 
with $D=1.5$ may not be residing in the nodes. Some of the galaxies which are embedded in curly filaments or are in vicinity of 
two or more filaments can also have $D=1.5$. The best way to judge that is to look at their local densities. We find that despite 
some of the flat galaxies having $D \sim 1.5$, very few of them have local density greater than the mean density. Hence we can 
infer that these flat galaxies either reside in curly filaments or they have more than one filaments inside the fitting radius. 
As a last remark we can say that this part of the analysis shows that flat galaxies are absent in cluster cores. Majority of the 
flat galaxies in our sample are found to be residing in low density regions like sheets and fields. In the top right panel of 
\autoref{fig:ldrn} we show the distribution of galaxies on the $D-{\cal R}_{25}$ plane. This plot shows more prominently the 
segregation of the flat and non-flat galaxies in terms of their local environment.\\

\noindent In the mid-panels of \autoref{fig:ldrn} the probability distribution function of $D$ and ${\cal R}_{25}$ is shown for 
the flat and non-flat galaxies. In both the figures, we find a clear distinction between the PDFs of flat and non-flat galaxies. 
The left panel shows that there are significant number of non-flat spirals closer to nodes or lie on sheet like structures, 
whereas the fraction of non-flat galaxies lying in under-dense regions like fields is comparatively much smaller. On the other 
hand, the flat galaxies are predominantly found in sheets and fields; a very small fraction of them have $D<1.5$. We find that 
only $19 \%$ of flat galaxies have $D \sim 1.5$ or less whereas more than $43 \%$ non spirals are found in dense regions like 
filaments or nodes. The median values of $D$ for flat and non-flat galaxies are $2.2$ and $1.8$ respectively. However, the third quantile 
for the flat galaxies is found to be $2.7$ which is much greater than the value $2.2$ for the non-flats. The respective median 
values of ${\cal R}_{25}$ for flat and non-flat galaxies are $12.9$ $h^{-1} {\rm Mpc}$ and $6.4$ $h^{-1} {\rm Mpc}$, which are 
well-separated as well. The PDF of ${\cal R}_{25}$ for these two classes of galaxies shown on the right panel clearly 
shows the difference in their distributions. ${\cal R}_{25}$ here serves as a proxy of $\eta$. Higher values of ${\cal R}_{25}$ indicates low 
density environments.\\

\noindent We carry out a two sample K-S test to quantify the differences in the probability distribution of $D$, $\eta$ and
${\cal R}_{25}$ for the two classes of galaxies. The bottom panels of \autoref{fig:ldrn} show the corresponding cumulative probability 
distributions same way as the mid-panels of the same figure. The green dotted lines indicate the values of $D$ and ${\cal R}_{25}$ 
where the supremum difference (\autoref{eq:Dks}) in the two cumulative distribution functions are found. This supremum differences 
(${\cal D}_{KS}$) are used to evaluate the K-S probability given in \autoref{eq:pks}. \\

\noindent \autoref{tab:1.1} tabulates the obtained results from the K-S tests. Comparing between the flat and the non-flat galaxies,
we find that the values of ${\cal D}_{KS}$ satisfies the criteria of $p<0.01$ for each of the distributions for $\eta$, $D$ and 
${\cal R}_{25}$. So it is quite clear that the majority of flat galaxies reside in an environment which is significantly different 
from the non-flat ones. The clear distinction in the distribution of local density for the flat and non-flat galaxies,
along with the difference in their local dimension indicates that the flat galaxies not only reside in a much 
rarer environment, also a majority of them lie in sheets or fields where the inflow of mass from the surrounding happens in very 
slow manner. This implies  that the flat galaxies mostly grow in isolation and rarely experience a major merger in their lifetime. 
That can be a main reason why these galaxies are mostly devoid of bulges because the bulge formation of these galaxies has to occur via secular evolution. Our study supports the findings of \citet{kormendy10} and \citet{grossi18} but does not comply with the models of formation of bulgeless galaxies by \cite{governato10}.

\subsection{Environment of flat galaxies and superthin flat galaxies from the group finding algorithm}

\noindent We next calculate the
${\kappa_{\rm{min}}}$ values for our sample of superthin flat and 
other flat galaxies. ${\kappa_{\rm{min}}}$ is a measure of the sparsity of the local environment; a higher $\kappa_{\rm{min}}$ value 
indicates a sparser local environment and vice-versa (\S 3.2).
The median $\kappa_{\rm{min}}$ values for all superthins 
and non-superthins in the sample are $2.3$ and $1.7$ respectively, indicating that both superthins and non-superthins are located 
in isolated environments in general, but the environment of the superthins is sparser compared to that of the non-superthins.\\


\noindent Next, we study the local environment of the galaxies in each of our sub-samples. The median values of $\kappa_{\rm{min}}$ of the \emph{isolated} ($\kappa_{\rm{min}}> 1$) superthins and \emph{isolated} non-superthins are $7.8$ and $6.6$ respectively. Similarly, the 
median $\kappa_{\rm{min}}$ values for the group ($\kappa_{\rm{min}}< 1$) superthins and non-superthins are $0.36 $ and $0.31$ 
respectively, in line with the earlier trend. ${\kappa_{\rm{comp}}}$ 
values for the superthins/non-superthins also show the same trend for the sub-samples except for the root-superthins. We find the median $\kappa_{\rm{min}}$ for the superthins are always higher compared to the non-superthins. Therefore, to summarize, \emph{superthins are located in sparser environments than non-superthins in general}. The number of superthins and non-superthins in different environments along with their median $a/b$ and median $\rm{{\kappa}_{min}}$ values have been summarized in \autoref{tab:2.1}. \\

\begin{figure*}
\resizebox{7.1 cm}{!}{\rotatebox{0}{\includegraphics{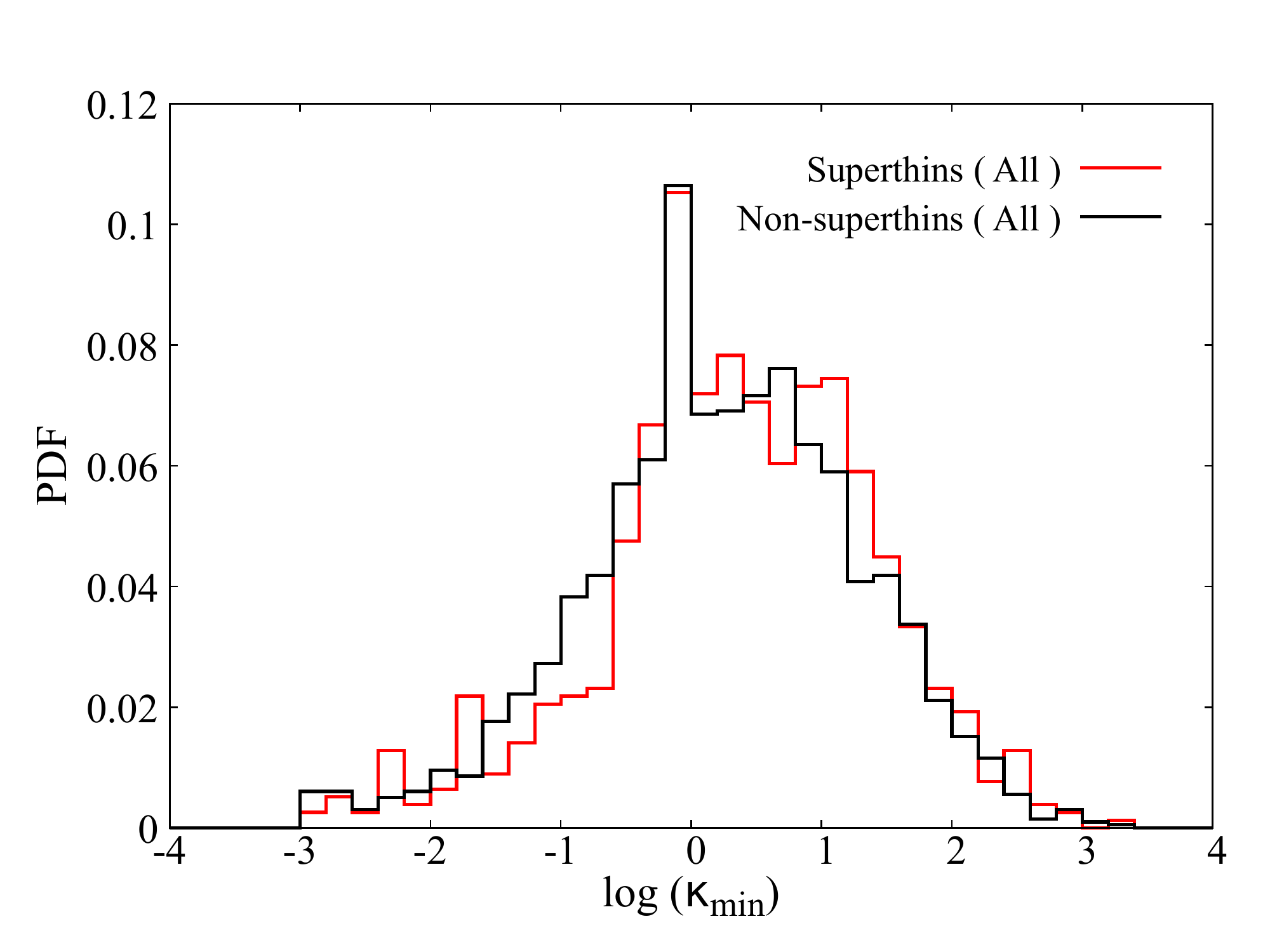}}} \hspace{0.3 cm}
\resizebox{7.1 cm}{!}{\rotatebox{0}{\includegraphics{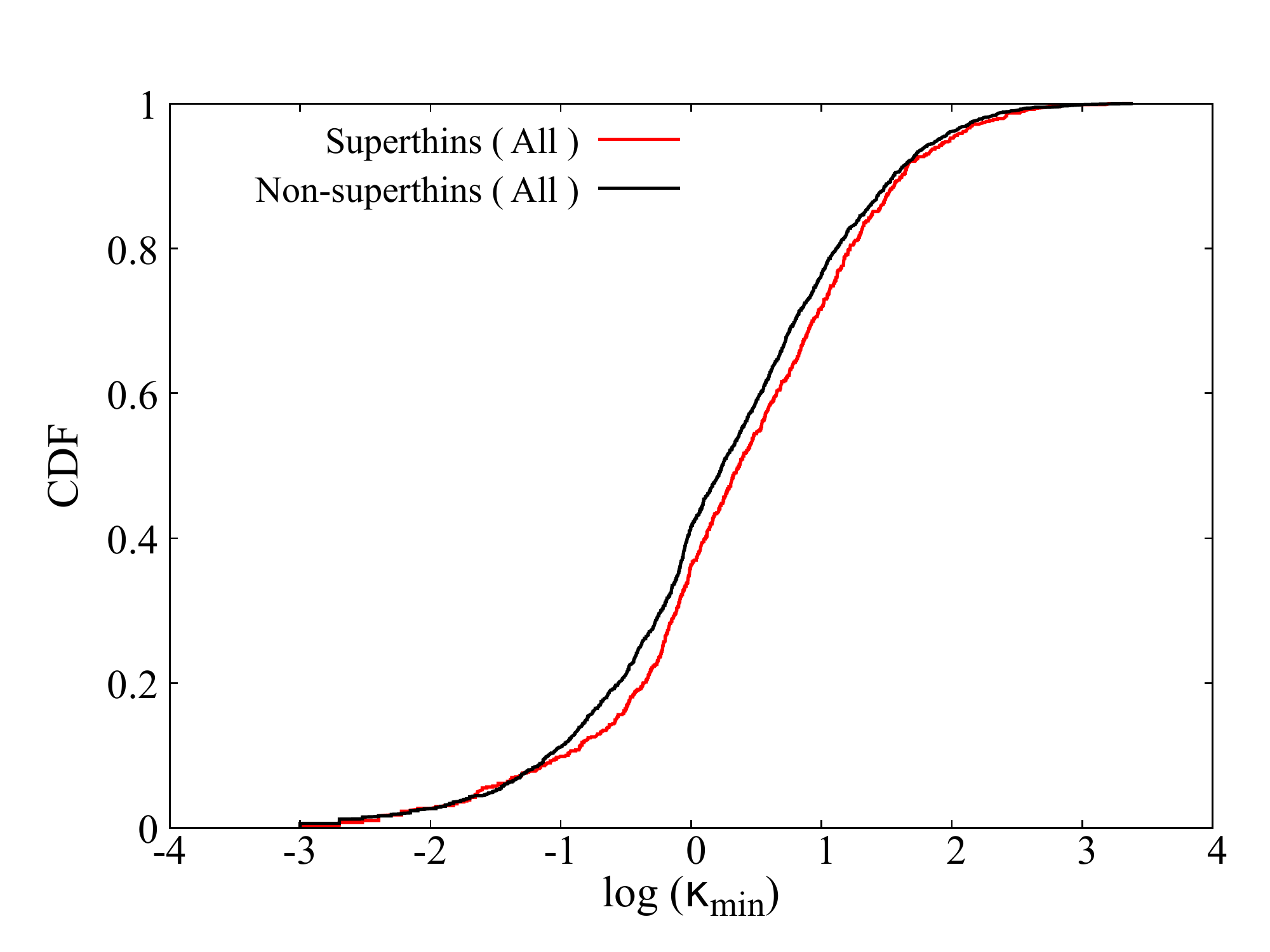}}} \\
\caption{Probability distribution function (PDF) and Cumulative distribution function (CDF) of the clusterization index $\kappa_{\rm{min}}$ indicating the density of the local environment in our sample for superthin and non-superthin flat galaxies.} 
\label{fig:2.1}
\end{figure*}

\noindent In \autoref{fig:2.1}, we present the probability distribution and cumulative distribution of ${\kappa_{\rm{min}}}$ (\S 3) for 
\emph{all (isolated + group)} superthin and non-superthin flat galaxies. We also carry out a 2-sample Kolmogorov-Smirnov (KS) test to 
check if the underlying populations of the superthin and the non-superthin galaxies in a given local environment significantly differ 
from each other. A $p$ value of $0.02$ obtained from the K-S test for ${\kappa_{\rm{min}}}$ indicates that the local environments of superthins and non-superthins are very different from each other. 
The same test is also done for $\kappa_{comp}$ to compare the different sub-classes of superthins and non-superthins.
For the isolated ones, the K-S test gives a $p$ value of $0.3$ implying that we cannot conclusively comment if the local environment of the \emph{isolated} superthins and \emph{isolated} non-superthins are significantly different from each other. We next compare the local environments of \emph{group} ($\kappa_{\rm{min}} < 1$) superthin and \emph{group} non-superthin galaxies. Again a $p$ value of $ 0.1 $ from K-S test indicates that the results are inconclusive. Finally we identify the galaxies in groups as {root} or {member}. We compare the environments of superthins and non-superthins in these two classes as well. 
We do not find any statistically significant difference in the $\kappa_{\rm{min}}$ in any of these two classes. Hence, we  only infer that \emph{the local environments of all (isolated + group) superthins and non-superthins seem to be significantly different from each other}. 
The K-S test performed for $\kappa_{comp}$ shows that there is no such noticeable difference between the superthins and non-superthins in 
terms of their presence near massive groups. All the results for K-S test are tabulated in \autoref{tab:2.2}. \\

\noindent Finally, we investigate the connection between the local environment and the disc vertical structure by obtaining the correlation 
between the clusterization index ${\kappa_{\rm{min}}}$ and the axes ratio $a/b$ for our various sub-samples. We note here that $a/b$ serves as the proxy for the planar-to-vertical axes ratios of the stellar discs of the galaxies. We find a low value of the 
correlation coefficient $r_{{\kappa_{\rm{min}}}}=0.03$ at low levels of statistical significance ($p = 0.15$) in general, implying 
almost no correlation of the disc axes ratio with the sparseness of the local environment. ${\kappa_{\rm{min}}}$ indicates the dynamical 
effect of the most immediate or important neighbour (host galaxy group) on an isolated or root (member) galaxy, which also is the most 
dominant effect. However, the influence of more distant galaxies (massive groups) in the local neighbourhood on the isolated galaxy 
(host galaxy group of a root or member galaxy, as a whole) may not be always negligible. We therefore calculate the correlation between 
${\kappa_{\rm{comp}}}$ and $a/b$ for our sub-samples. We again find a low correlation coefficient r$_{{\kappa_{\rm{comp}}}} = -0.0003$ at 
high levels of statistical significance ($p = 0.99$) again implying no correlation at all. We further investigate the correlation between 
the axes ratio and clusterization indices separately in superthin and non-superthin galaxies and found no significant correlation in any of 
the two types of samples. Therefore, we conclude, that \emph{disc axes ratio of both superthin and non-superthin galaxies are almost 
uncorrelated with the sparsity of the local environment.} \\


\begin{table*}{}
\caption{Results: Local environment of superthin and non-superthin flat galaxies in different environments using the group finding algorithm}
\label{tab:2.1}
\begin{tabular}{lcccc}
\hline
Sub-sample  & $N_{eff}$ &  Mean $a/b$ & Median ${\kappa}_{min}$ & Median ${\kappa}_{comp}$\\
\hline
All Superthins & $779$ & $11.8 \pm 2.04$ & $2.3$ & $8.2$ \\
All Non-Superthins & $1983$ & $8.25 \pm 0.8$ & $1.7$ & $7.1$ \\
& & \\
Isolated Superthins & $496$ & $11.9 \pm 2.2$ & $7.8$ & $10.4$ \\
Isolated Non-superthins & $1157$ & $8.28 \pm 0.8$ & $6.5$  & $9.1$ \\
& & \\
Group Superthins & $283$ & $11.68 \pm 1.9$ & $0.36$  & $5.5$ \\
Group Non-Superthins & $826$ & $ 8.2 \pm 0.8$ &  $0.31$  & $5.3$ \\
& & \\
Root Superthins & $72$ & $11.6 \pm 1.9$ & $0.2$  & $5.1$   \\
Root Non-Superthins  & $275$ & $8.16 \pm 0.8$ & $0.17$ & $6.7$  \\
& & \\
Member Superthins & $211$ & $11.7 \pm 1.9$ & $0.4$ & $5.9$ \\
Member Non-Superthins & $551$ & $8.22 \pm 0.8$ & $0.38$  & $4.6$ \\
\hline
\end{tabular}
\begin{itemize}
\item[] \scriptsize{$N_{eff}$ is the effective number count, $a/b$ the major-to-minor axes ratio. $1-\sigma$ standard deviation around the mean is also shown. ${\kappa}_{min}$ and ${\kappa}_{comp}$ are the clusterization indices.}
\end{itemize}
\end{table*}

\begin{table*}{} 
\caption{Results: K-S test for the clusterization indices of superthins and non-superthins in different local environments from the group finding algorithm} 
\label{tab:2.2} 
\begin{tabular}{lcccc} 
\hline 
Sub-sample &$D_{KS}$ ($\kappa_{min}$) & $p$( $\kappa_{min}$) & $D_{KS}$ ($\kappa_{min}$) & $p$($\kappa_{comp}$)  \\ 
\hline 
All & 0.06  & 0.02 & 0.04 & 0.3 \\ 
&&&& \\ 
Isolated & 0.05 & 0.3 & 0.05 & 0.4  \\ 
&&&& \\ 
Group & 0.07 & 0.14 & 0.05 & 0.7 \\ 
&&&& \\  
Root & 0.1 & 0.4 & 0.1 & 0.3 \\ 
&&&& \\ 
Member & 0.06 & 0.6 &  0.06 & 0.7 \\ 
\hline 
\end{tabular}
\begin {itemize}
\item[] \scriptsize{$\cal{D}_{KS}$ is the supremum distance in K-S test, and $p$ is the corresponding probability.  ${\kappa}_{min}$ and ${\kappa}_{comp}$ are the clusterization indices.}
\end{itemize}
\end{table*}

\section{Discussion}
\noindent \textbf{Effect of redshift space distortion:} It is a well-accepted fact that the calculation of local dimension is partly affected by redshift space distortions (RSD). On small scales, some spurious filament-like structures may appear due to {\it Finger of God} effect, elongating the virial clusters in redshift space. A proper modelling for the effect of RSD is required to address that issue. However, the effect of RSD is dominant on small scales and here we have chosen $n=25$, which corresponds to ${\cal R}_{25} \sim 8 \hmpc$ on average. The value of $R_2$ being amplified by the exponential term would always be higher than ${\cal R}_{25}$. Galaxies lying in denser regions 
like clusters would still be affected. \citet{pandey20} shows that the effect of RSD in measuring local dimension is quite small for 
$D \sim 1.5$ and persists up to $\sim 25 \hmpc$. So we have a small window where RSD would partly affect the analysis. As our goal here 
is not to identify filaments and clusters or to discriminate them, the results shown here despite of RSD carry sufficient evidence of 
the flat galaxies to have a clearly distinct environment compared the other spirals.\\

\noindent \textbf{Local dimension of superthin flat galaxies:} Due to the lack of enough number of data points, we were unable to carry out the analysis involving local dimension $D$ with the superthin flat galaxies alone. Using the criteria $a/b >10 $ to identify the superthins, we were left with only $78$ 
galaxies in the given region. This number is quite small to conduct a statistical study and infer anything conclusive. Nevertheless we carried out an analysis to check for any difference in the probability distributions of $D$ for superthins and non-superthins. However,
a K-S test on the $D$ distributions of the superthin and non-superthin population gives a $p$ value $0.2$ for $D$. Similarly, for the local density $\eta$, we get $p=0.98$. Therefore, we do not find any conclusive evidence of the superthins to have a different local geometrical environment compared to the other flat galaxies. Hence, we used
the group finding technique to separately study the environmental differences between the superthins and other flat galaxies in RFGC.\\

\noindent \textbf{Local environment of ultra-superthins: } We repeat our analysis of group finding with identifying the superthins using two 
more choices of critical axes ratio ($a/b=12$ \& $a/b=15$) and found out that the median $\kappa_{min}$ for the superthins ($3.01$ and $4.5$ 
respectively) are again higher compared to the non-superthins ($1.8$ and $1.9$) in each of the cases. The $p$ value of K-S test for ${\kappa_{\rm{min}}}$ 
monotonically reduces to $p=0.014$ and $p=0.010$ respectively for these two choices. Also the difference in the environment of group superthins turns out to be statistically significant in those cases with $p=0.05$ and $0.07$. {\it This might hold the clue to the origin of the small $p$-value obtained in the K-S test of $\kappa_{{\rm{min}}}$ for all (isolated + group) superthins and non-superthins.} Hence we conclude that although the numerical values of the results 
from the group finding algorithm are sensitive to the choice of the values of the axes ratio used to define the superthins, the general trend remains unchanged.

\section{Conclusion}
 We compare the local environment of flat galaxies in the revised flat galaxy catalogue (RFGC) \citep{kara99} with the non-flat galaxies 
as characterized in the Galaxy Zoo project. We prepare a three dimensional hybrid map combining the flat galaxies from RFGC, non-flat galaxies from Galaxy Zoo and galaxies from SDSS spectroscopic 
data, and investigate the local geometrical environment of each of the flat galaxies using the local dimension and local density estimators. 
Here we try to quantify the type of a geometric structure each of the flat galaxies reside in. A subset of the flat galaxies are superthins
which have a strikingly large major-to-minor axes ratio $a/b > 10$. Our results shows that the flat galaxies, including super thins, are absent in relatively high density regions 
like galaxy cores. A large fraction of the flat galaxies is found to have local density way below the mean 
density and have a local dimension of $D>2$ which implies that these galaxies grow in isolation and rarely experience a major merger in their 
lifetime. Hence, the bulge formation of these galaxies has to be via a secular evolution. Our analysis shows that the flat galaxies 
including the superthins have a distinctly different environment compared to the non-flat galaxies at a confidence level of more than $99\%$. \\

\noindent The second part of our analysis aims to study empirically if the existence of the superthin flat galaxies is linked to the density of their local environment. We choose a sample of galaxies from RFGC and determine if they are located 
in a isolated/group environment by using the available data on the line-of-sight velocities and the projected distances between the galaxies. We find that superthins and non-superthin flat galaxies are found to be located in different local environments, at a  $98\%$ statistical significance, which may have 
important implications for their formation and evolution models. However, we find that the major-to-minor axes ratios of the stellar discs, which serves as the proxies of the planar-to-vertical axes ratios,  are almost uncorrelated with the densities of the local environment, for of both superthin and non-superthin galaxies. This possibly indicates that environment does not play a dominant role in regulating the vertical 
thickness of the stellar discs in late-type, flat galaxies.\\

\section{Data availability}
Data used in this paper are publicly available at SDSS and RFGC Databases. The data produced through this work will be shared on reasonable request 
to the corresponding author.

\section{ACKNOWLEDGEMENT}  
DM acknowledge the support by the Russian Science Foundation, grant 19-12-00145. 
Funding for the Sloan Digital Sky Survey IV has been provided by the Alfred P. Sloan Foundation, the U.S. 
Department of Energy Office of Science, and the Participating Institutions. SDSS-IV acknowledges support and 
resources from the Center for High Performance Computing  at the University of Utah. The SDSS website is www.sdss.org.
SDSS-IV is managed by the Astrophysical Research Consortium for the Participating Institutions of the SDSS 
Collaboration including the Brazilian Participation Group, the Carnegie Institution for Science, Carnegie 
Mellon University, Center for Astrophysics | Harvard \& Smithsonian, the Chilean 
Participation Group, the French Participation Group, Instituto de Astrof\'isica de Canarias, The Johns 
Hopkins University, Kavli Institute for the Physics and Mathematics of the Universe (IPMU) / University 
of Tokyo, the Korean Participation Group, Lawrence Berkeley National Laboratory, Leibniz Institut f\"ur 
Astrophysik Potsdam (AIP),  Max-Planck-Institut f\"ur Astronomie (MPIA Heidelberg), Max-Planck-Institut 
f\"ur Astrophysik (MPA Garching), Max-Planck-Institut f\"ur Extraterrestrische Physik (MPE), National 
Astronomical Observatories of China, New Mexico State University, New York University, University of Notre 
Dame, Observat\'ario Nacional / MCTI, The Ohio State University, Pennsylvania State University, Shanghai 
Astronomical Observatory, United Kingdom Participation Group, Universidad Nacional Aut\'onoma de M\'exico, 
University of Arizona, University of Colorado Boulder, University of Oxford, University of Portsmouth, 
University of Utah, University of Virginia, University of Washington, University of Wisconsin, Vanderbilt 
University, and Yale University. RFGC uses data collected from the Digitized Sky Survey, which was produced 
at the Space Telescope Science Institute under U.S. Government grant NAG W-2166.

\bsp	
\label{lastpage}
\end{document}